\documentclass[12pt]{iopart}

\usepackage{iopams}  
\usepackage{graphicx}
\graphicspath{ {./Figures/} }
\usepackage{float}
\expandafter\let\csname equation*\endcsname\relax
\expandafter\let\csname endequation*\endcsname\relax
\usepackage{amsmath}
\usepackage{multirow}
\usepackage{hhline}
\usepackage{color}
\usepackage[pdftex,colorlinks=true,linkcolor=blue,citecolor=blue,urlcolor=blue]{hyperref}
\usepackage{centernot}
\usepackage{mathtools}
\usepackage{stmaryrd}
\usepackage{lineno}
\usepackage{csquotes}
\MakeOuterQuote{"}
\begin{document}

\title[GEO\,600 beam splitter thermal compensation system: new design and commissioning]{GEO\,600 beam splitter thermal compensation system: new design and commissioning}

\author{S\'everin Nadji$^{1,*}$, Holger Wittel$^1$, Nikhil Mukund$^{2,3}$, James Lough$^1$, Christoph Affeldt$^1$, Fabio Bergamin$^1$, Marc Brinkmann$^1$, Volker Kringel$^1$, Harald Lück$^1$, Michael Weinert$^1$ and Karsten Danzmann$^1$}

\address{$^1$ Max Planck Institute for Gravitational Physics and University of Hannover, D-30167 Hannover, Germany}
\address{$^2$ LIGO, Massachusetts Institute of Technology, Cambridge, MA 02139, USA}
\address{$^3$ NSF AI Institute of Artificial Intelligence and Fundamental Interactions, Massachusetts Institute of Technology, Cambridge, MA 02139, USA}

\ead{severin.nadji@aei.mpg.de}
\vspace{10pt}
\begin{indented}
\item[]\today
\end{indented}

\begin{abstract}
Gravitational waves have revolutionised the field of astronomy by providing scientists with a new way to observe the universe and gain a better understanding of exotic objects like black holes. Several large-scale laser interferometric gravitational wave detectors (GWDs) have been constructed worldwide, with a focus on achieving the best sensitivity possible. However, in order for a detector to operate at its intended sensitivity, its optics must be free from imperfections such as thermal lensing effects. In the GEO\,600 gravitational wave detector, the beam splitter (BS) experiences a significant thermal lensing effect due to the high power build-up in the Power Recycling Cavity (PRC) combined with a very small beam waist. This causes the fundamental mode to be converted into higher order modes (HOMs), subsequently impacting the detector's performance. To address this issue, the GEO\,600 detector is equipped with a thermal compensation system (TCS) applied to the BS. This involves projecting a spatially tunable heating pattern through an optical system onto the beam splitter. The main objective of the TCS is to counteract the thermal lens at the BS and restore the detector to its ideal operating condition. This paper presents the new beam splitter TCS in GEO\,600, its commissioning, and its effect on strain sensitivity. It also outlines the planned upgrade to further enhance the performance of the TCS.
\end{abstract}


\section{Introduction}
\label{Intro}
GEO\,600 \cite{luck2010upgrade} is a gravitational wave (GW) detector in Hannover, Germany. It is made of 600 m long folded arms in a dual recycling cavity configuration \cite{grote2004dual}. Ideally, when a gravitational wave passes through, it modulates the TEM$_{00}$ carrier light signal measured at the dark port of the detector using the DC readout scheme \cite{hild2009dc}. However, the dark port of the detector contains signals other than the fundamental mode. These signals include sidebands that control the different optical cavities in the detector and unwanted higher order modes (HOMs) caused by misalignment \cite{mukund2023neural,mukund2020bilinear} or imperfections in the different optics used. Nevertheless, only the fundamental mode which potentially carries the gravitational wave signal, is transmitted through an Output Mode Cleaner (OMC) to the final photodiode which used to compute the strain sensitivity of the detector \cite{hewitson2003calibration}.
 
\begin{figure}[H]
\centering
\includegraphics[width=\textwidth]{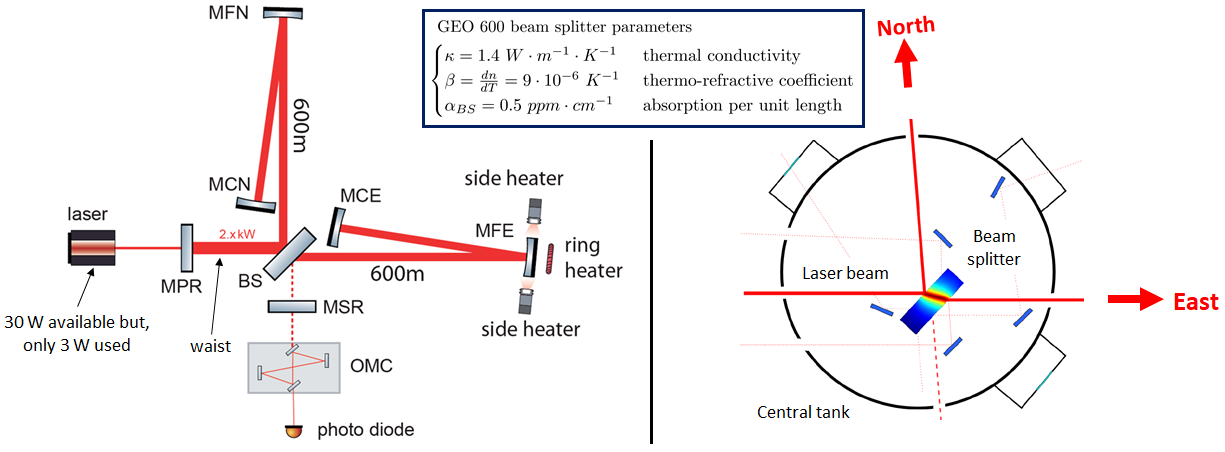}
\caption{GEO\,600 layout and the thermal profile through the beam splitter \cite{wittel2009compensation}.}
\label{fig:1}
\end{figure}

At high frequencies, the sensitivity of GEO\,600 is limited by shot noise due to photon counting statistics. One way to alleviate that limitation is by injecting squeezed light through the dark port of the interferometer. This technique has been implemented at GEO\,600 and has shown to reduce the shot noise by over 6dB \cite{GEO6dB}. Another solution is increasing the interferometer's circulating power ($\text{P}_{\text{cir}}$) \cite{affeldt2014laser}. To do so, in GEO\,600, we use a Power Recycling Cavity (PRC) \cite{drever1983gravitational,schnier1997power}, which allows a power build-up that can nearly reach a factor of 800. However, given the architecture of the GEO\,600, a high circulating power introduces a strong thermal lensing effect in the beam splitter (BS). In fact, GEO600 does not have arm cavities, and the PRC is formed by the detector operating in the dark fringe offset configuration and the power recycling mirror, as shown on the left in Figure \ref{fig:1}. Therefore, half of the power built up in the PRC passes through the fused silica BS and is partially absorbed with an absorption coefficient of $0.5~\text{ppm}\cdot \text{cm}^{-1}$ at $\lambda=1064~\text{nm}$. Thus, the wavefront propagating in the east arm is modified and mismatches with the one in the north. Consequently, the fundamental mode gets converted into unwanted HOMs, leading to a degradation of the interferometer contrast and even reduced power build-up in the PRC, more generally impacting detector performance \cite{brooks2007hartmann}. Moreover, the HOMs can also introduce spurious signals to auxiliary photodetectors used to align the detector subsystems and optical cavities \cite{schreiber2016alignment}. 

As a gravitational wave detector, GEO\,600 has been at the forefront of using thermal compensation systems based on the principles of blackbody radiation. The first application involved the use of a ring heater and two side heaters to adjust the radius of curvature of the East Folded Mirror (MFE) \cite{luck2004thermal,wittel2014thermal}, as shown on the left in  Figure \ref{fig:1}. This adjustment partially compensates for the east arm beam mismatch caused by the thermal lens in the beam splitter. However, pushing the compensation too far would result in an optically unstable PRC \cite{wittel2015active}. Therefore, unlike the other GW detectors, in GEO\,600, we directly compensate for the thermal lensing effect at the BS by projecting through an optical system a spatially tunable heating pattern generated with a Heater Matrix (HM) \cite{wittel2018matrix}. The heat reaching the beam splitter is then absorbed and locally changes the refractive index, hence the wavefront, by the thermo-refractive effect. As for the thermal expansion and thermo-elastic effects, they are relatively low and will be neglected as shown in \cite{wittel2009compensation}. It is important to note that the third generation of gravitational wave detectors, namely the Einstein Telescope High Frequency (ET-HF) \cite{abernathy2011einstein} and the Cosmic Explorer (CE) \cite{galaxies10040090}, will operate with kilowatt scale laser power at their beam splitters. Hence, a strong thermal lensing effect can be expected, as for GEO\,600.
 
This paper will be divided into three sections. In section 2, we will provide a detailed description of the new beam splitter thermal compensation system implemented in GEO\,600. Section 3 will present experimental results that will help us evaluate the TCS's performance. In Section 4 we will summarise and give the future prospects of this work before concluding.

%___________________________________________________________________
%___________________________________________________________________
\section{Description and mitigation of the BS thermal lensing effect in GEO\,600}
%-----------------------------------------------------------

\subsection{The beam splitter thermal lensing effect in GEO\,600}
\label{HOMsVsPower}
\begin{figure}[H]
\centering
\includegraphics[width=1\textwidth]{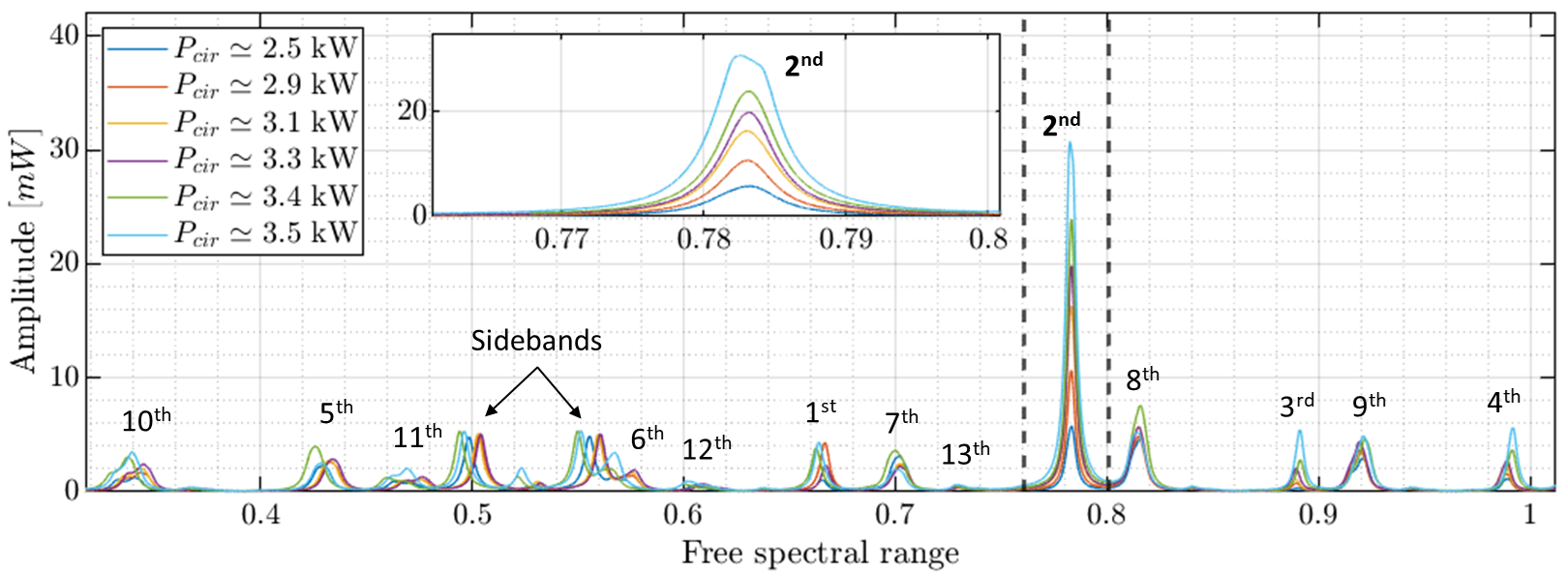}
\caption{OMC scans for different circulating powers.}
\label{fig:2}
\end{figure}

As mentioned in section \ref{Intro}, one way to lower the shot noise limited sensitivity at high frequencies is to increase the circulating power in GEO\,600. However, this approach has its drawbacks. Higher circulating power leads to a more substantial thermal lensing effect in the beam splitter. We can measure the  strength of the thermal lensing effect by monitoring the power fluctuation in the dark port. Additionally, conducting a mode scan with the OMC (as shown in Figure \ref{fig:2}) can help identify the main consequences of the thermal lensing effect.

These mode scans indicate that the mainly 2$^\text{nd}$ HOM gets more substantial along with some even HOMs as the operating power increases, which is a symptom of mode mismatch of the main beam to GEO\,600. Indeed, when the laser propagates through the BS, it gets partly absorbed and creates a lens-like media. At steady state, the ABCD matrix of such lens-like media is given by \cite{xu2006propagation}:

\begin{equation}
\begin{pmatrix}
\text{A} & \text{B}\\
\text{C} & \text{D}
\end{pmatrix}
=
\begin{pmatrix}
\cos\left(\gamma_{\text{x},\text{y}} \text{z}_{\text{max}}\right) & \sin\left(\gamma_{\text{x},\text{y}} \text{z}_{\text{max}}\right)/\gamma_{\text{x},\text{y}}\\
\\
-\gamma_{\text{x},\text{y}}\sin\left(\gamma_{\text{x},\text{y}} \text{z}_{\text{max}}\right) & \cos\left(\gamma_{\text{x},\text{y}} \text{z}_{\text{max}}\right)
\end{pmatrix}
\label{eq:2}
\end{equation}
with $\gamma_{\text{x},\text{y}}$ the created medium coefficient along the x and y directions, and z$_{\text{max}}$ the beam pathlength in the BS. By using the generalized Fresnel transform, we can express the lens-like media ABCD matrix as a product of different ABCD matrices that define simple known optical systems such as Propagation (P), Magnification (M), and Lens transformation (L) \cite{palma1997extension}. Therefore, for $\gamma_{\text{x},\text{y}}\text{z}_{\text{max}}\ll 1$, we can rewrite the thermal lens defined in Equation \ref{eq:2} and show that:

\begin{equation}
\text{P}
\simeq  
\begin{pmatrix}
1 & \text{z}_{\text{max}}\\
0 & 1
\end{pmatrix}\quad
\text{M}
\simeq 
\begin{pmatrix}
1 & 0\\
0 & 1
\end{pmatrix}\quad
\text{L}
\simeq 
\begin{pmatrix}
1 & 0\\
-\gamma_{\text{x},\text{y}}^2\text{z}_{\text{max}} & 1
\end{pmatrix}
\end{equation}

\noindent
Hence, GEO\,600 operating at high circulating power is analogous to its original design, with an additional asymmetric lens exclusively in the east arm. The asymmetric profile of the resulting thermal lens is because the laser beam crosses the BS at an angle of approximately $41^{\circ}$, causing the beam radius on the x-axis $w_\text{x}$ to differ from that on the y-axis $w_\text{y}$. The term $\gamma_{\text{x},\text{y}}^2\text{z}_{\text{max}}$ known as the refractive power of the lens L of focal length f$_{\text{therm}-\text{x},\text{y}}$, is given by \cite{strain1994thermal,bogan2015novel}:
\begin{equation}
\gamma_{\text{x},\text{y}}^2\text{z}_{\text{max}} = \frac{1}{\text{f}_{\text{therm}-\text{x},\text{y}}} = 1.3\frac{\beta}{2\kappa\pi}\frac{\text{P}_{\text{cir}}}{w^2_{\text{x},\text{y}}}\alpha_{\text{BS}}z_{\text{max}}
\label{eq:3}
\end{equation}

In GEO\,600, the waist $w_0\simeq 10~$mm is located at the BS, which has a thickness of $\text{d}_{\text{BS}} = 80~\text{mm}$. At a nominal circulating power $\text{P}_{\text{cir}}\simeq 2.9~\text{kW}$, the focal lengths of the created thermal lenses in the x and y axis are respectively 8.95 km and 6.14 km. In ET-HF \cite{abernathy2011einstein} and CE \cite{galaxies10040090}, the targeted power in the arm cavities are respectively $\text{P}_{\text{arm, ET-HF}}\simeq 3~\text{MW}$ and $\text{P}_{\text{arm, CE}}\simeq 1.5~\text{MW}$. While their optical designs are still in progress, we expect at least $\text{P}_{\text{cir, ET-HF}}=\text{P}_{\text{cir, CE}}\simeq 10~\text{kW}$ in their power recycling cavities. Although the absorption level of bulk fused silica has been reduced to $\alpha=0.25~\text{ppm}\cdot \text{cm}^{-1}$ \cite{punturo2010third}, the created thermal lenses at the BS will still strongly depend on the beam radius, which should ideally be larger than in GEO\,600.

%-----------------------------------------------------------
\subsection{The beam splitter thermal compensation method in GEO\,600}

In GEO\,600, we use a heater matrix (HM) \cite{wittel2015active,wittel2018matrix} composed of $9\times 12$ heater elements (PT100) that can be individually addressed. This allows us to generate different heating patterns, which are then projected onto the beam splitter (BS) through an optical system that we will introduce shortly. This thermal compensation method has two main benefits:

\begin{itemize}

\item Low noise coupling into the strain sensitivity: Actuating directly on the beam splitter to compensate for the thermal lensing effect would most likely affect the optical path length of the main laser beam, so couple into displacement noise. On the one hand, we have CO$_{2}$ laser-based TCS, which requires an active intensity stabilization to keep the relative intensity noise (RIN) relatively low \cite{ballmer2006ligo}. On the other hand, we have black body radiation-based TCS, which, contrarily to a CO$_\text{2}$ laser, produces uncorrelated photons and presents a much lower RIN \cite{lawrence2003active}. For instance, if the heating power absorbed at the BS is $\text{P}_\text{abs}=7.5~\text{W}$, the resulting RIN would be 500 times lower than that of a well-stabilized CO2 laser ($\text{RIN}=10^{-7}$). A detailed comparison for GEO\,600 can be found in \cite{wittel2009compensation}, where it was also demonstrated that the thermal compensation system (TCS) induced noise is approximately 10 times smaller than the design sensitivity of GEO-HF at 1 kHz.

\item Compensation pattern adaptability: The heater matrix provides a high degree of flexibility when selecting a heating pattern. Our initial focus was optimizing the heating pattern to minimize the 2$^\text{nd}$ HOM content, as detailed in \cite{wittel2015active}. As a result, we identified the so-called $\text{HG}_{\text{20}}$ heating pattern that will be exclusively used in this paper. The same approach can be repeated, and one can define a set of $\text{HG}_{\text{nm}}$ heating patterns to reduce the $(\text{n}+\text{m})$ HOM content, and then apply the principle of superposition to reduce simultaneously different HOM contents with a combination of patterns.

\end{itemize}

\section{Hardware of the TCS in GEO\,600}
The original TCS in GEO600 had two main limitations. First, the HM printed control board (PCB) burned due to the high heat generated. Second, the optical system had a small numerical aperture \cite{wittel2015active}, which limited the amount of heat reaching the beam splitter (BS). Therefore, our upgrades were mainly focused on developing a TCS with better heat insulation for the HM and a tunable optical system to allow better heat transfer towards the BS.

%-----------------------------------------------------------
\subsection{The heater matrix}

\begin{figure}[H]
\centering
\includegraphics[width=0.9\textwidth]{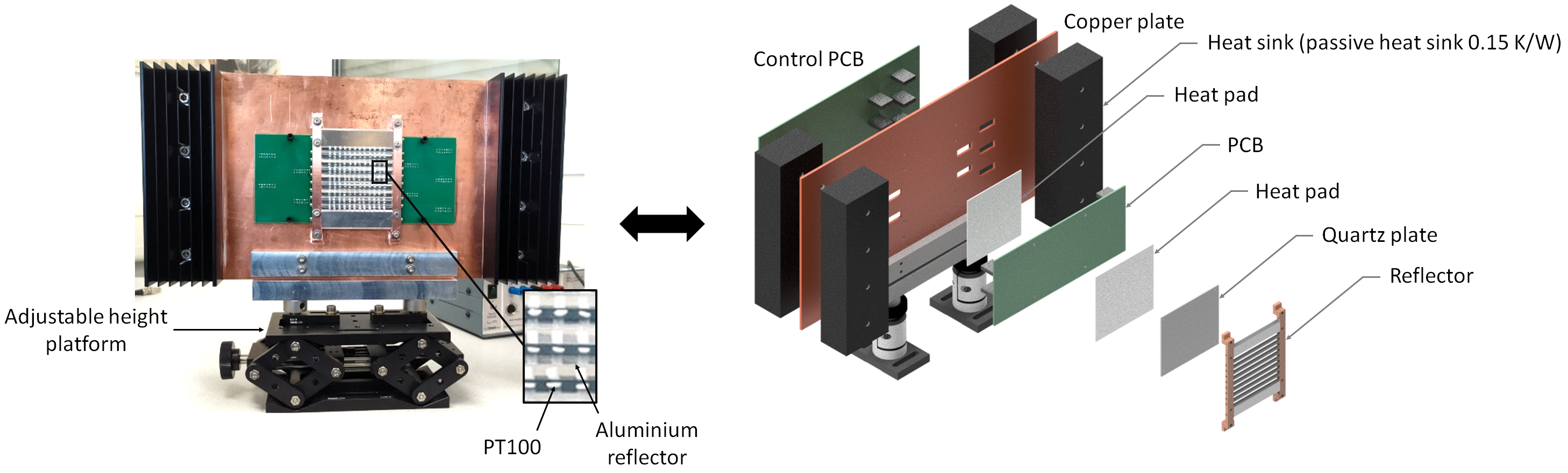}
\caption{Photograph of the new HM assembly, and its schematic exploded view.}
\label{fig:3}
\end{figure}

We designed a new HM assembly with improved heat dissipation capabilities to address the first limitation mentioned above. The main purpose was to create a way out for the unused heat generated by the heater elements, which flows back to the PCB. To achieve this, we have placed a first heat pad between a quartz plate (isolating the PT100 elements +/-pins) and the front side of the PCB to contain the generated heat. A second heat pad is placed between a copper plate and the back side of the PCB to efficiently circulate the heat towards the copper plate on which four heat sinks are mounted. Each provides a passive cooling of $0.15~\text{K}\cdot \text{W}^{-1}$, which ensures the continuous dissipation of unused heat and prevents the PCB from burning. We have also designed the reflector holders in copper, establishing a connection to the main copper plate to ensure that the absorbed heat gets dissipated in the same manner as described above. In our current configuration, the heater elements' maximum operating temperature is estimated to be T$_{\text{max}}=873~^\circ \text{K}$. Figure \ref{fig:3} shows a picture of our new HM assembly on the left side and a 3D model of its exploded view (the heater elements not represented) on the right side. Our new assembly has proven to be very robust, allowing us to operate at the highest generated heat for years without any observed degradation.

%-----------------------------------------------------------
\subsection{The optical system}

\begin{figure}[H]
\centering
\includegraphics[width=0.9\textwidth]{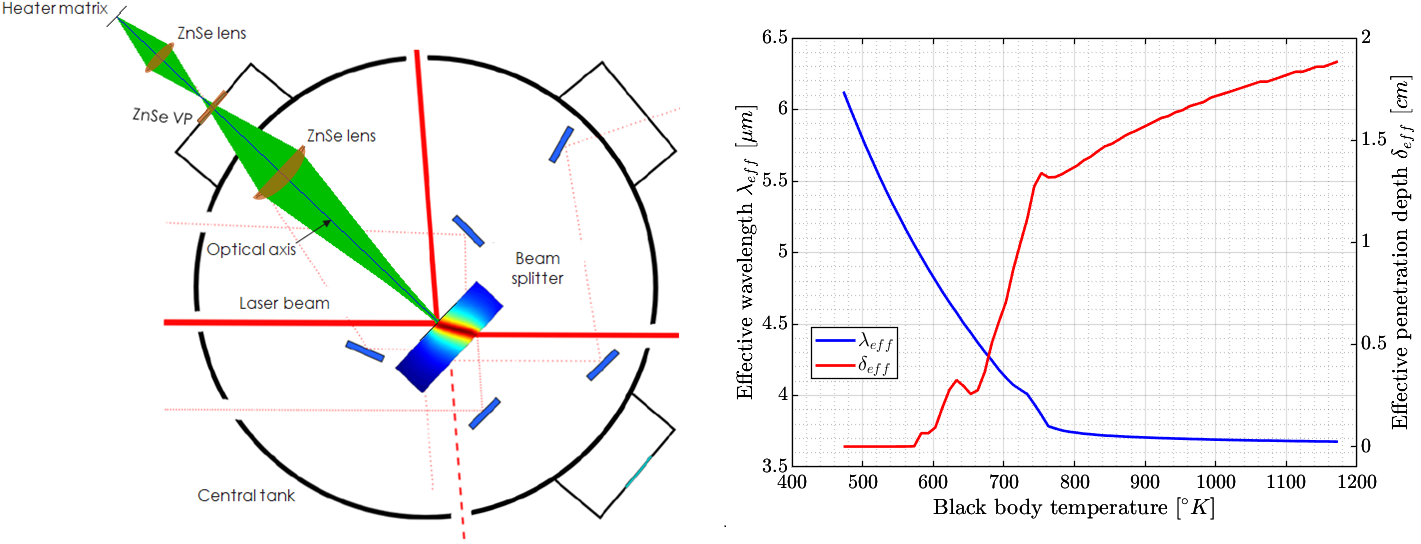}
\caption{On the left is a simplified picture of the thermal compensation system in GEO\,600. On the right side, the temperature dependence of the effective radiated wavelength $\lambda_{\text{eff}}$, and the corresponding penetration depth $\delta_{\text{eff}}$.}
\label{fig:4}
\end{figure}

Once generated, the heating pattern is sent to the beam splitter using an optical system that includes two ZnSe lenses placed in a series. The first lens, which is located outside the central tank, has a diameter of d$_1=80~$mm and a focal length of f$_1 = 63.75~$mm at $\lambda=7.5~\mu$m. Its purpose is to image the radiated pattern in front of the central tank's viewport. The second lens, with a diameter d$_2=160~$mm and a focal length of f$_2 = 170~$mm at $\lambda=7.5~\mu$m, is placed in vacuum in the central tank to project the final image of the heating pattern onto the beam splitter. Figure \ref{fig:4} depicts a schematic of the new TCS implemented in GEO\,600 on the left side. It features a numerical aperture (NA) of $0.15$ compared to $0.06$ for the initial TCS design. 

According to the Beer-Lambert Law, one can express the total absorbed power per wavelength along the $\text{z}$ axis through the BS as follows:

\begin{equation}
\text{P}_{\text{abs}}(\lambda,\text{T},\text{z})={\cal A}_{\text{BB}}\cdot \text{M}(\lambda,\text{T})\cdot\left(1-e^{-\alpha_{\text{BS}}(\lambda) \text{z}}\right)
\end{equation} 

\noindent 
with ${\cal A}_{\text{BB}}$ representing the total surface area of the black body, $\text{M}(\lambda,\text{T})$ the Plank radiation from the heating pattern, and $\alpha_{\text{BS}}(\lambda)$ the spectral absorption of the beam splitter. Hence, the effective wavelength $\lambda_{\text{eff}}$ of the optical system is chosen to maximise P$_{\text{abs}}(\lambda,\text{T},\text{z})$, and the corresponding penetration depth is given by:

\begin{equation}
\delta_{\text{eff}}(\lambda)=\frac{1}{\alpha_{\text{BS}}(\lambda)} 
\end{equation} 
\noindent 
Figure \ref{fig:4} shows on the right side the temperature dependence of the effective wavelength in blue and the corresponding penetration depth in red. The lower the temperature, the longer the wavelength and all the radiated heat is absorbed at the surface of the beam splitter.

\noindent 
On the other hand, the higher the temperature, the shorter the wavelength, eventually reaching an asymptote at $\lambda^{\text{lim}}_{\text{eff}}\simeq 3.7~\mu$m corresponding to a penetration depth $\delta^{\text{lim}}_{\text{eff}}\simeq 1.8~$cm. Therefore, we can precisely adjust the magnification and the focus of the projected heating pattern for $\lambda^{\text{lim}}_{\text{eff}}$ (PT100 heating at T$\simeq 873~^\circ$K) by adapting the HM and the ZnSe lens positions derived from the lens transformation equation:
\[
\text{HM} \xrightarrow[]{\text{f}_1(\lambda_\text{T})} IM_{\text{HM},1} \xrightarrow[]{\text{f}_2(\lambda_\text{T})} IM_{\text{HM},2}
\]
where $IM_{\text{HM},1}$ and $IM_{\text{HM},2}$ denote the images of the heating pattern through, respectively, the first and second ZnSe lenses. In addition, we will characterise the TCS as defocused by $x~$mm ($\text{DEF }x$) when the heating pattern is imaged at $x~$mm from the beam splitter plane. For a given magnification factor, this will allow the projected heating pattern to be more or less spread on the beam splitter.

%-----------------------------------------------------------
\section{Experimental results}
%-----------------------------------------------------------
In this section, the experiment was carried out with the detector locked in DC mode. This locking scheme allows the computation of the strain sensitivity of GEO\,600, which is needed to assess the performances of the TCS further.

A straightforward optimisation approach for the TCS optical system would consist of defining the optimal magnification and defocus of the projected heating pattern by appropriately adjusting the longitudinal positions of the HM and the outer ZnSe$_1$ lens. Therefore, the optimisation happens in two steps: adjusting the magnification and then the defocus to minimise the overall dark port power before the OMC. In this section, we present the experimental results obtained and evaluate the performance of our new TCS.

%-----------------------------------------------------------
\subsection{TCS optimisation for GEO\,600 nominal operating power}
\begin{figure}[H]
\centering
\includegraphics[width=\textwidth]{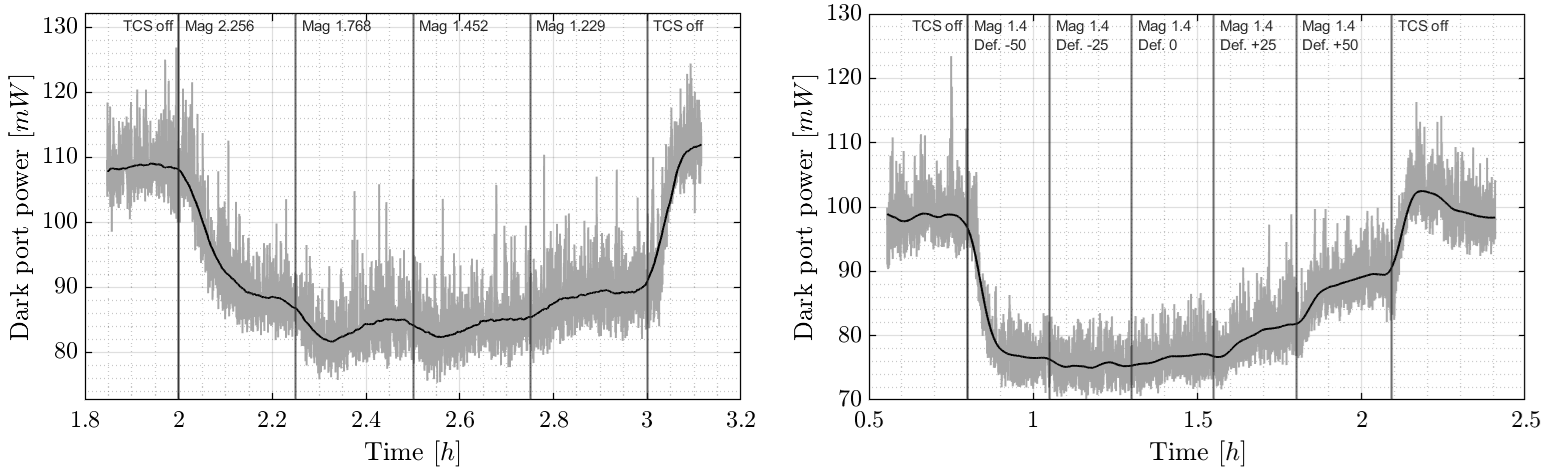}
\caption{DPP for different Mag (left), and for different Def at a selected Mag (right).}
\label{fig:5}
\end{figure}

The optimisation work was performed for a circulating power of $\text{P}_{\text{cir}}\simeq$ 2.9 kW without the squeezing operating. Figure \ref{fig:5} shows the GEO\,600 dark port power variation during the process. On the left, the impact of different magnification factors (Mag) is shown, and on the right, the defocus tuning (Def) for the selected magnification factor. During the magnification tuning, the dark port power reaches a global minimum before increasing again. We can also see that for $\text{Mag}~1.77$ and $\text{Mag}~1.45$, the dark port power (DPP) decreases to a lower value before increasing slightly again. This suggests that the TCS performed slightly better in the transition between different magnification factors, which is akin to a defocused state. From the optimisation process, we have determined that $\left( \text{Mag}~1.4, \text{Def}~-25 \right)$ is the optimal TCS configuration for the HG$_\text{20}$ heating pattern.

\begin{figure}[H]
\centering
\includegraphics[width=0.8\textwidth]{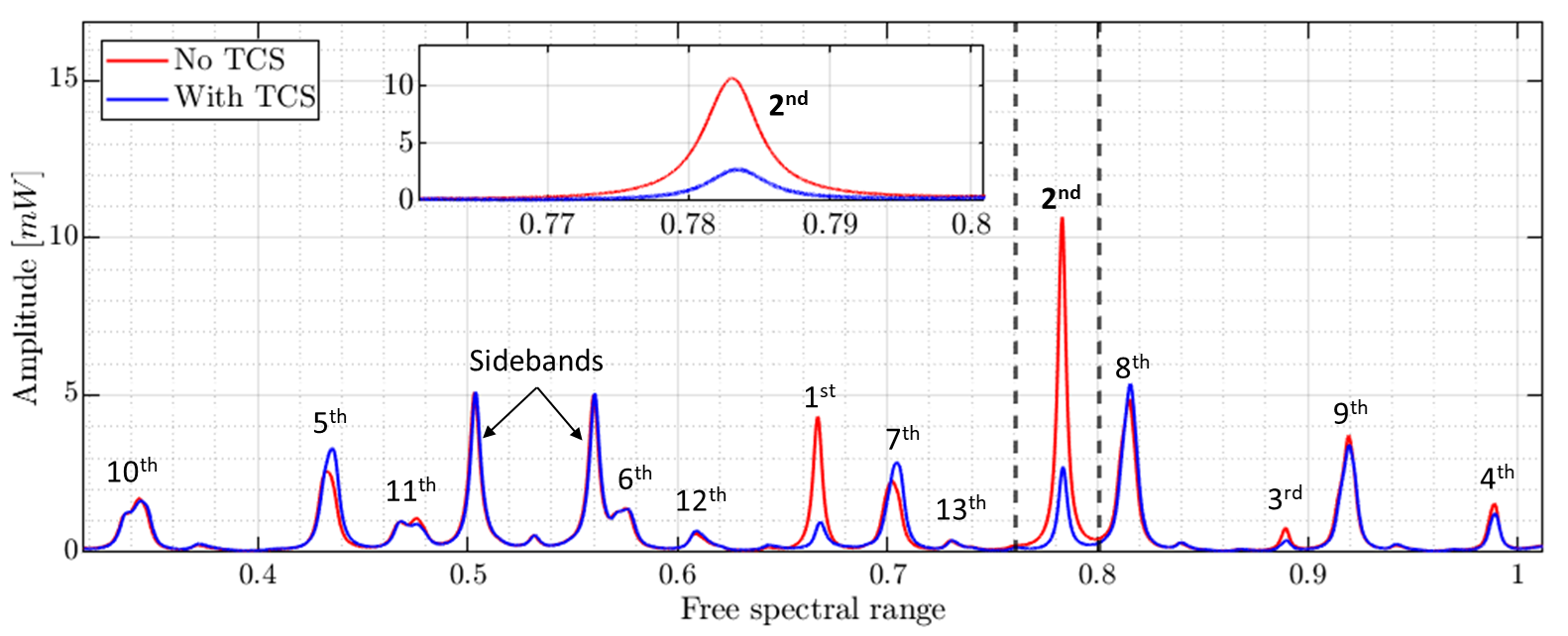}
\caption{OMC scan with and without TCS for a circulating power $\text{P}_{\text{cir}}=$ 2.9 kW.}
\label{fig:6}
\end{figure}

It would also be interesting to analyse the individual HOMs in the dark port of the GEO\,600 with the optimised TCS operating. To do so, we locked the detector in heterodyne mode \cite{hild2009dc} so that we could perform an OMC scan. The results without and with TCS are depicted in figure \ref{fig:6} in red and blue, respectively. 

\noindent 
In our observations, the second higher order mode (HOM) decreased drastically by about 70\%. The other even HOMs appeared to be relatively stable in both cases. This provides strong evidence that using the TCS on the beam splitter can significantly improve the matching of modes between the north and east arms of the detector. As for the odd HOMs, it is essential to note that the OMC auto-alignment loops are disengaged in heterodyne mode. Therefore, any misalignment we observe during the OMC scan (with or without TCS) should, in principle, be corrected when the detector is locked in DC mode.

%-----------------------------------------------------------
\subsection{Effect of the optimised TCS on GEO\,600 at nominal operating power}
\label{Effect_tcs_p29}

\begin{figure}[H]
\centering
\includegraphics[width=\textwidth]{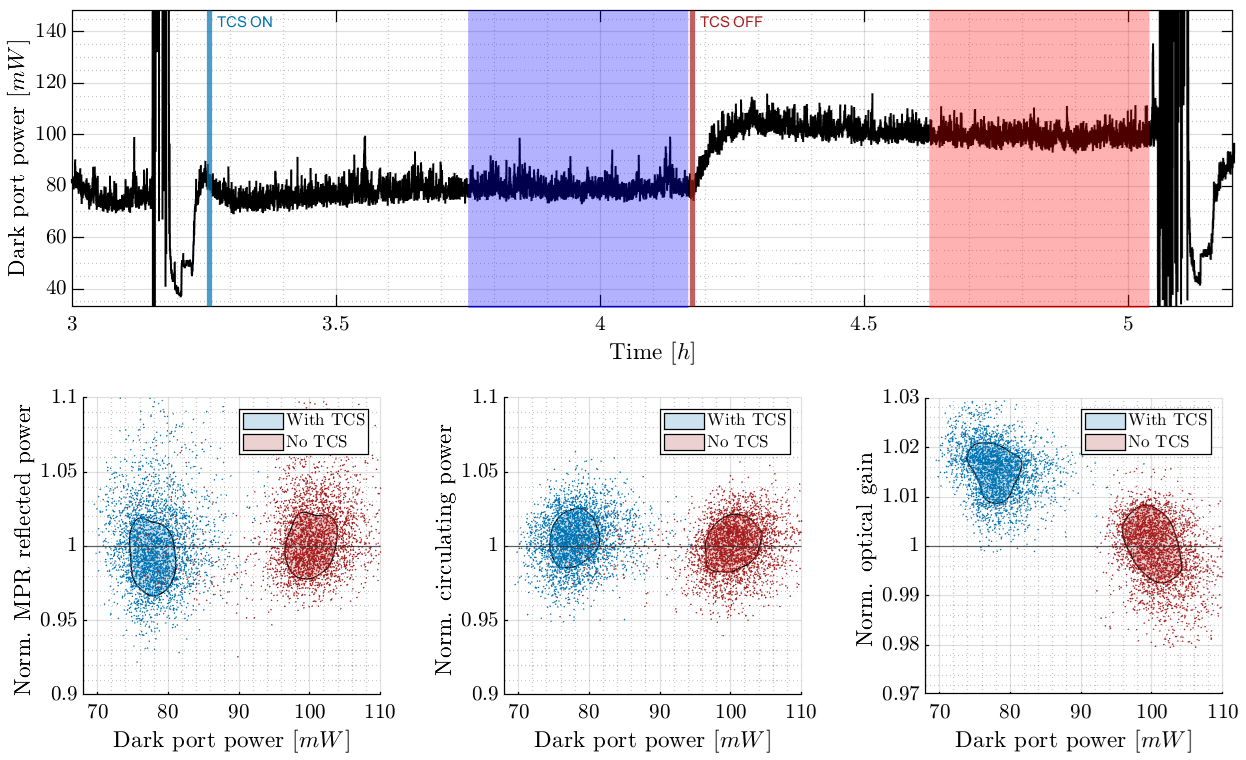}
\caption{At the top is the dark port power time series. Below are the scatter plots of key parameters versus dark port power. The black contours represent the lower ($16^\text{th}$) and upper ($84^\text{th}$) percentiles for the bound on the density estimate.}
\label{fig:7}
\end{figure}

To demonstrate the impact of the TCS on GEO\,600, we compared DC-locked sessions with and without TCS. In Figure \ref{fig:7}, the first raw depicts the dark port power time series for these two sessions over an extended period during which the detector remained in the same state without any lock loss or significant seismic level changes. The only difference between the two sessions was the presence or absence of TCS. The blue vertical line in the graph indicates the activation of TCS when the detector was locked. Approximately an hour later, we switched off the TCS, as shown by the red vertical line. 

\begin{figure}[H]
\centering
\includegraphics[width=\textwidth]{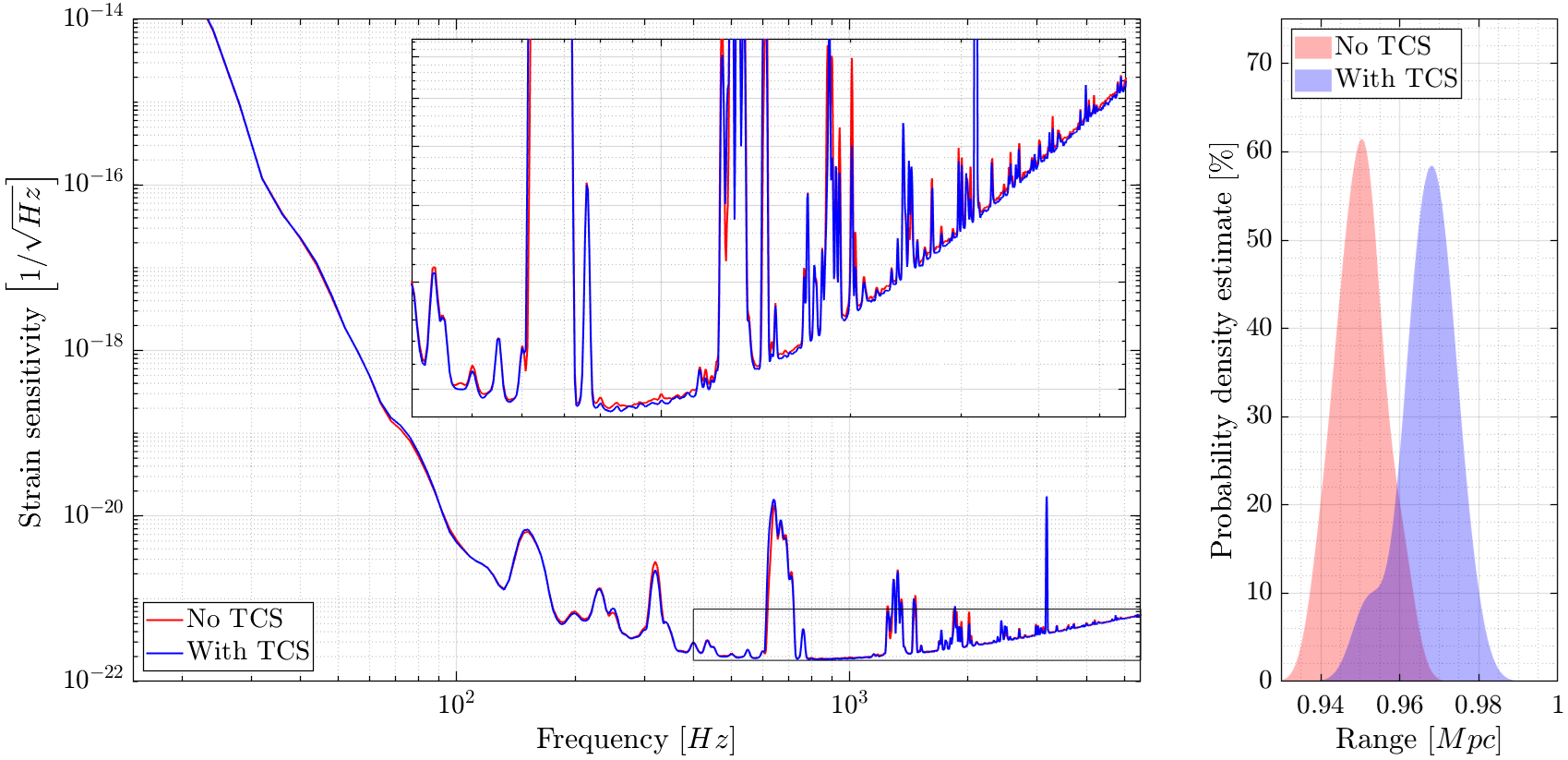}
\caption{KDE of the observe range without (red) and with (blue) TCS.}
\label{fig:8}
\end{figure}

As we anticipated, the dark port power is lower when the TCS operates. Additionally, other parameters also infer the performance of the TCS, including the reflected power of the MPR, the effective circulating power, and the optical gain of GEO\,600. In Figure \ref{fig:7}, the second row displays the 2D kernel density estimation (KDE) of the three aforementioned parameters normalized to the case where the TCS was not operating. The data without and with TCS are shown in dark red and blue, respectively. When the TCS was in operation, we observed a reduction in reflected power from the MPR, leading to an increase in circulating power and improved optical gain. The pole frequency of the GEO\,600 optical plant, which was constant in both cases, is not shown here. We have also compared the strain sensitivity of the GEO\,600 with and without TCS, and the results can be seen on the right side in Figure \ref{fig:8}. In addition, we analysed the improvements in detection range for a binary neutron star system of 1.4 solar masses each, with a signal-to-noise ratio of 8. Our calculations started at a minimum frequency of 50 Hz and terminated at the frequency that marks the innermost stable circular orbit (FISCO): $\text{FISCO} = 4.4~\text{kHz}\cdot (\text{M}_1+\text{M}_2)^{-1} = 1571~\text{Hz}$. We computed the detection range every two minutes within the two states highlighted in blue and red in Figure \ref{fig:7}. On the right side of Figure \ref{fig:8}, we have the corresponding KDE. With the TCS operating, we observed an increase in the detection range, corresponding to the observed optical gain improvement.

\noindent

It is crucial to understand that the primary purpose of the TCS is to mitigate the thermal lensing effect at the beam splitter, thus helping the detector recover its intended state. Therefore, improvement in the detection range should be interpreted as an improvement towards the detector's designed sensitivity at $\text{P}_{\text{cir}}$=2.9 kW. Besides, whether the TCS is active or not, the KDE bandwidths remain consistent in both scenarios. This suggests and confirms that the TCS does not add extra noise to the strain sensitivity, which we can see in the strain sensitivity plots.

%-----------------------------------------------------------
\subsection{Optimised TCS for different circulating power}
\begin{figure}[H]
\centering
\includegraphics[width=\textwidth]{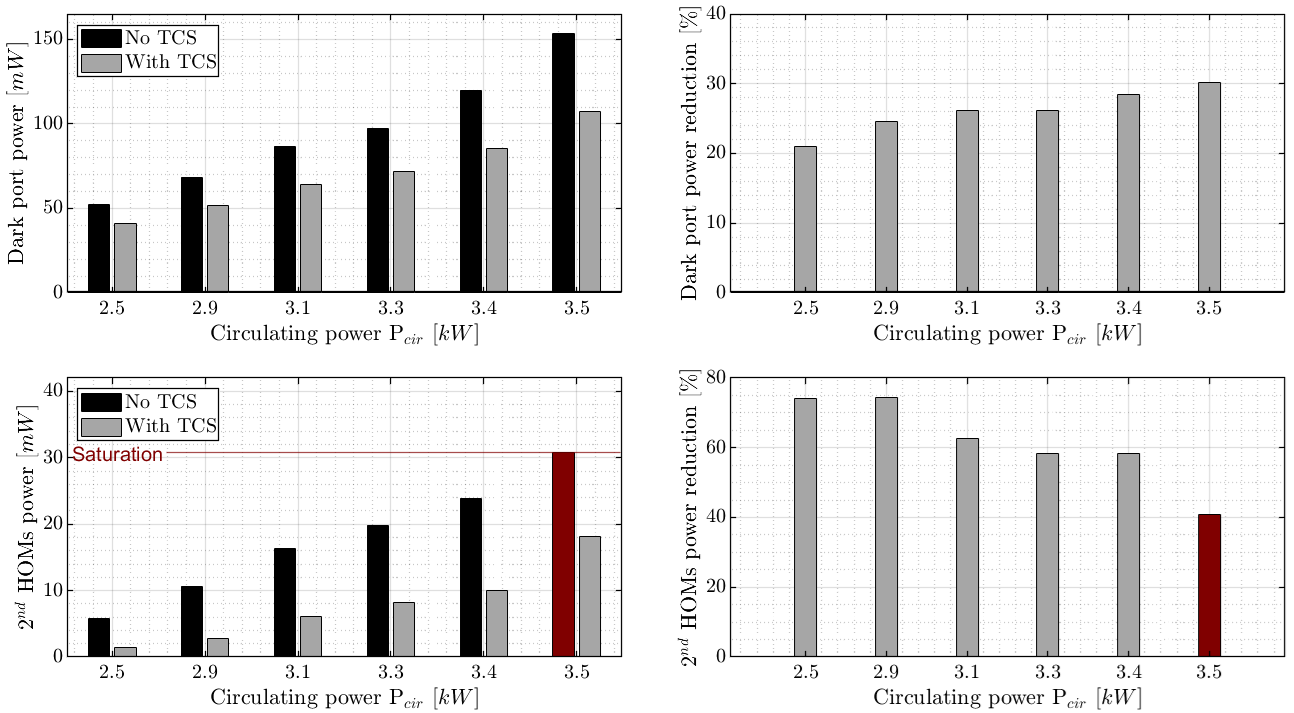}
\caption{For different $\text{P}_{\text{cir}}$, the DPP with and without TCS (top left), and the corresponding reduction rate (top right). The $\text{2}^{\text{nd}}$ HOM power with and without TCS (bottom left) and the corresponding reduction rate (bottom right).}
\label{fig:9}
\end{figure}

Once an optimal configuration was determined, we increased the circulating power in GEO\,600 to see how well the TCS performed at higher power levels. During this experiment, the detector was locked in heterodyne mode \cite{hild2009dc}, so without the dark fringe offset, as it is more stable at higher circulating powers. Furthermore, It is important to stress that the optimisation process is most likely power-dependent. However, the goal here is to determine the least HOMs reduction we can achieve for higher circulating power and, therefore, a more substantial thermal lensing effect in the beam splitter when the TCS is well tuned for the nominal operating power of GEO\,600. Figure \ref{fig:9} shows the dark port power levels without and with TCS for different circulating power and the corresponding reduction rates in the first row. The second row shows the second HOM content fluctuation, similarly without and with TCS for different circulating power, and the corresponding reduction rates. It is worth pointing out that the performance of the tuned TCS scales slightly with the operating power.
Nevertheless, we have observed a limit at high circulating powers, likely due to the limited available radiated power. Increasing the TCS radiated power throughput would be logical as the thermal lensing effect grows. Importantly, we have achieved good performance with the chosen TCS configuration, enabling a 50\% to 70\% reduction in the second HOM content. Furthermore, it is important to stress that the current TCS allows the detector to operate at higher power while keeping the HOMs as low as in the nominal operating power. For instance, operating at $\text{P}_{\text{cir}}$=3.3 kW with the TCS is equivalent to operating at $\text{P}_{\text{cir}}$=2.9 kW without TCS, which has been so far our default configuration.

%-----------------------------------------------------------
\section{Summary and Outlook}

We have developed and implemented a new thermal compensation system to counteract the thermal lensing effect at the GEO\,600 beam splitter. Our new heater matrix assembly design, with improved thermal insulation, allows us to operate safely and continuously at the highest temperature without damaging its printed control board. Furthermore, the upgraded optical system, featuring a larger numerical aperture and essential additional capabilities such as magnification and defocus tuning, ensures optimal heat distribution of the projected pattern on the beam splitter.

Once optimised at nominal circulating power $\text{P}_{\text{cir}}$=2.9 kW, we assessed the performance of the TCS by comparing the GEO\,600 data with and without TCS. Our analysis revealed an improvement in the optical gain of the detector, therefore the detection range for a binary neutron star system, as depicted in Section \ref{Effect_tcs_p29}. It is crucial to note that we did not observe any additional noise affecting the strain sensitivity during the TCS operation. This indicates that the noise introduced by the TCS is below the current GEO\,600 sensitivity curve. At higher circulating power $\text{P}_{\text{cir}}$=3.5 kW, the new TCS also performed well, allowing up to $30$\% reduction of unwanted higher-order modes.

As part of the ongoing upgrade, we will replace the current PT100 elements with a different type of heater called PT12, which will have an approximatively $40\%$ larger surface area for the same surface material and can reach up to $\text{T}_{\text{max}}=973~^\circ \text{K}$.

\begin{figure}[H]
\centering
\includegraphics[width=0.9\textwidth]{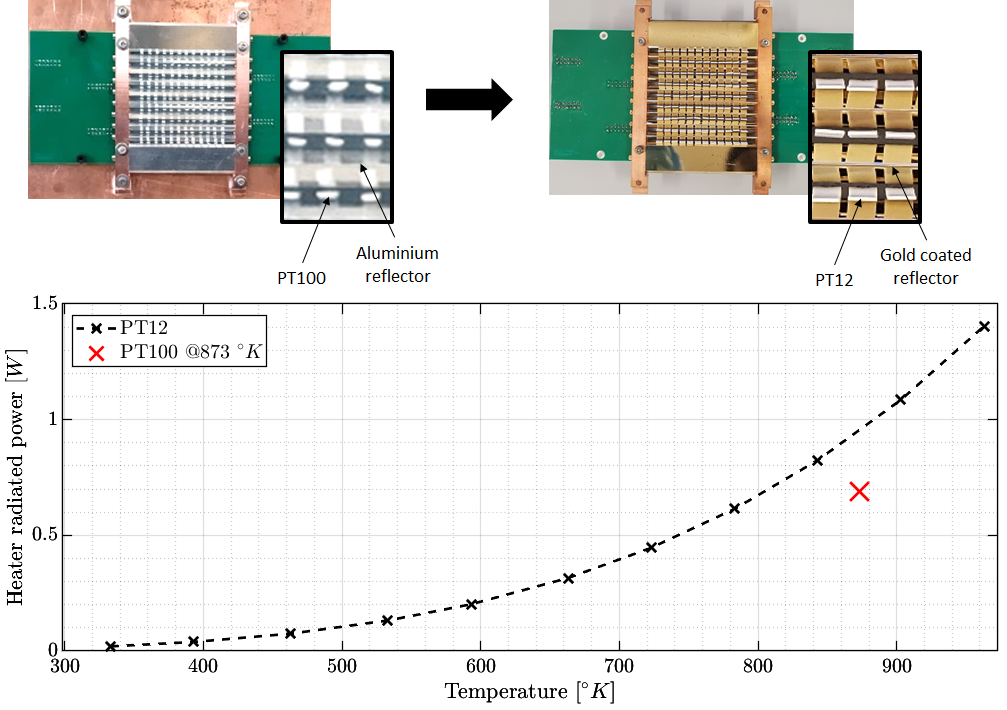}
\caption{Temperature dependence of the PT100 and PT12 radiated power.}
\label{fig:10}
\end{figure}

\noindent
As per the well-known Stefan Boltzmann law, we have computed and compared the thermally radiated power between the current heater type (represented by red dots) and a PT12 heater (represented by a dashed black line) as shown in Figure \ref{fig:10}. Additionally, we have replaced the aluminium reflector with a gold-coated one to better reflect heat towards the optical system. With all these new features, we expect a more effective mitigation of the thermal lensing effect for the same TCS configuration. 

In our future plans, we aim to increase the operating power of GEO\,600. To achieve this, we will iteratively optimise the TCS each time we reach a stable operating point during the power-up process. It is crucial to note that reducing the HOMs is necessary as we increase the circulating power, but more steps are needed to ensure a stable locked interferometer. Several control loops are involved in keeping the GEO\,600 locked, which will require some tuning. Our efforts will allow us to understand better the challenges associated with the power-up process and mitigate them effectively. This is essential for commissioning the current and third-generation gravitational wave detectors, namely, ET-HF and CE, where the thermal lensing effect at their beam splitters is expected to be especially strong.
%___________________________________________________________________
\section{Acknowledgments}
The authors would like to thank Walter Grass for his  years of expert support in the maintenance of critical infrastructure to the site to include the extensive 400 m$^3$  vacuum system. The authors are grateful for support from  the Science and Technology Facilities Council Grant Ref: ST/L000946/1, the Bundesministerium für Bildung und Forschung, the state of Lower Saxony in Germany, the Max Planck Society, Leibniz Universität Hannover, and Deutsche Forschungsgemeinschaft (DFG, German Research Foundation) under Germany’s Excellence Strategy—EXC 2123 QuantumFrontiers—390837967. This document has been assigned LIGO document number LIGO P2400212.

\section*{References}
\bibliographystyle{iopart-num}
\nocite{TitlesOn}
\bibliography{sample}
%%%%%%%%%%%%%%%%%%%% APPENDIX%%%%%%%%%%%%%%%%%%%55

\end{document}